\documentclass[review, conference]{IEEEtran}
\IEEEoverridecommandlockouts
\usepackage{cite}
\usepackage{amsmath,amssymb,amsfonts}
\usepackage{algorithmic}
\usepackage{graphicx}
\usepackage{textcomp}
\usepackage{xcolor}
\def\BibTeX{{\rm B\kern-.05em{\sc i\kern-.025em b}\kern-.08em
    T\kern-.1667em\lower.7ex\hbox{E}\kern-.125emX}}

\usepackage{mypackage}
\usepackage{soul}

\begin{document}

\title{{\workname}: Exploring Migration on Virtualized CGRAs\\
\thanks{ICCS National Technical University of Athens \\
European Union, Project HIGHER}
}

\author{
\IEEEauthorblockN{
Agamemnon Kyriazis \qquad
Panagiotis Miliadis \qquad
Dimitris Theodoropoulos \vspace{.3ex} \\
\vspace{.3ex}
Nectarios Koziris \qquad
Dionisios Pnevmatikatos
}
\IEEEauthorblockA{Computing Systems Laboratory - National Technical University, Athens, Greece}
\vspace{.3ex}
\IEEEauthorblockA{akyriazis@cslab.ece.ntua.gr}
}

\maketitle

\begin{abstract}
As modern Coarse Grain Reconfigurable Arrays (CGRAs) grow in size, efficient utilization of the available fabric by a single application becomes increasingly difficult. Existing CGRA mappers either fail to utilize the available fabric or rely on rigid static code transformations with limited adaptability. Multi-tenant CGRAs have emerged as a promising solution to increase hardware utilization, but current attempts fail to address key challenges such as fabric fragmentation and live migration. To address this gap, we present \workname, an end-to-end system for CGRA multi-tenancy that supports dynamic scheduling and resource allocation in a shared environment. \workname\ addresses fabric fragmentation caused by kernels completing out of order by supporting both stateless and stateful live kernel migration as a de-fragmentation mechanism. We assess our solution on an Alveo-U280 data-center-grade FPGA card, reporting area, frequency, and power. Performance is evaluated using routines from the PolyBench benchmark suite and kernels derived from common machine learning operators. Results show that spatial sharing of the available fabric across multiple users improves workload makespan by up to $70.48\%$, while live kernel migration reduces tail latency on fragmented layouts by up to $29.60\%$. The custom tightly coupled controller and read-back paths required for virtualization and stateful migration introduce a LUT cost of 0.13\% per region. Our evaluation reveals that multi-tenancy is important for efficient CGRA utilization, and live kernel migration can further improve performance by recovering fragmented space with minimal hardware cost.
\end{abstract}


\section{Introduction}
Coarse Grain Reconfigurable Arrays (CGRAs) have emerged as a promising alternative to Field Programmable Gate Arrays (FPGAs), as they offer high post-fabrication flexibility and provide near ASIC energy efficiency~\cite{10.1145/3357375}. These features are, especially appealing for modern HPC systems and AI~\cite{sambanova, iea2025energyai}. Computation in CGRAs is performed both spatially and temporally, encouraging data re-use and minimizing memory transactions. Additionally, CGRAs are software programmable at a coarser granularity, resulting in lower reconfiguration overhead than FPGAs.

At the same time, utilizing the fabric of a CGRA efficiently at scale~\cite{2502.19114v2, 9835572} remains challenging for most existing DFG-based compilers. \autoref{tab:Compiler Performance} shows the minimum initiation interval achieved in several applications by increasing the grid size in spatiotemporal architectures.  Increasing the CGRA grid size beyond a certain point yields no additional benefit unless further optimizations are applied, leaving parts of the array idle despite available resources. Kernel fine-tuning can increase throughput, but statically elaborating such optimizations still requires manual effort. \autoref{tab:Mapper Performance} shows that similar issues are also observed in spatial architectures. For example, GEMM and 2MM use less than 27\% and 45\% of the available PEs, respectively, on an \(8\times8\) array. As a result, the amount of fabric area a single application can occupy on its own is limited without fine-tuned compilation.


\begin{table}[t]
  \centering
  \caption{Initiation Interval achieved with SAT-MapIt modulo scheduler~\cite{10.1145/3587135.3591433} for varying CGRA grid sizes. Minimum II achieved is marked in bold and underlined\protect\footnotemark[1].}
  \label{tab:Compiler Performance}
  {\small
  \setlength{\tabcolsep}{4pt}
  \begin{tabular}{c | c c c c}
    \hline
    Benchmark & 2x2 & 3x3 & 4x4 & 5x5 \\
    \hline
    GEMM    & 4   & \underline{\textbf{3}}  & 3             & 3 \\
    GEMM-u2 & 6   & \underline{\textbf{3}}  & 3             & 3 \\
    VADD    & 4   & \underline{\textbf{3}}  & 3             & 3 \\
    VADD-u2 & 9   & 4           & \underline{\textbf{3}}    & 3 \\
    \hline
  \end{tabular}
  }
  \vspace{-0.5ex}
\end{table}

\begin{table}[t]
  \centering
  \caption{Utilization achieved with GenMap spatial mapper \cite{genmap/9149647} for varying kernel unroll factors for an \(8\times8\) array\protect\footnotemark[2].}
  \label{tab:Mapper Performance}
  {\small
  \setlength{\tabcolsep}{3pt}
  \begin{tabular}{c | c c | c c | c c }
    \hline
    Benchmark & \multicolumn{2}{| c |}{-no-u} & \multicolumn{2}{| c |}{-u2} & \multicolumn{2}{| c |}{-u4} \\
    \hline
    Metric & bbox & exact & bbox & exact & bbox & exact \\
    \hline
    GEMM    & 6.35\%  & 6.35\%  & 15.87\% & 11.11\%  & 77.78\%  & 26.98\% \\
    2MM     & 7.94\%  & 7.94\%  & 22.22\% & 17.46\%  & 88.89\%  & 44.44\% \\
    SAXPY   & 6.35\%  & 3.17\%  & 9.52\%  & 6.35\%   & 88.89\%  & 30.16\% \\
    \hline
  \end{tabular}
  }
  \vspace{-2.5ex}
\end{table}
\footnotetext[1]{Benchmarks labeled “-u2” feature a $\times2$ unroll on the inner loop.}
\footnotetext[2]{BBox: smallest rectangle that contains all mapped operations. Exact: actual PE cell count that are occupied.}

These observations have motivated recent efforts to multiplex multiple applications on the same CGRA fabric. This idea is not new in reconfigurable computing. Virtualizing and sharing FPGA resources have been extensively studied as a way to improve hardware utilization, reduce response time, and provide system-level services~\cite{osdi18-khawaja, coyote/10.5555/3488766.3488822, coyotev2/10.1145/3731569.3764845, vital/0.1145/3373376.3378491, pmiliad/10.1145/3648475, nyx/10.1145/3695053.3731094, pipearch/10.1145/3418465}. Inspired by this, several works~\cite{PPA/10.1145/1669112.1669160, kong2023hardware, drips/9773269, multisky/11121915, cadas/10.1145/3793672} have explored whether sharing CGRAs resources across multiple workloads could emerge as an alternative to fully utilize the available fabric. This question has become increasingly relevant as CGRAs have evolved beyond their traditional edge-scale design, often sized at \(4\times4\) to \(8\times8\) PE grids~\cite{openedgecgra/10.1145/3587135.3591437, picachu/10.1145/3676641.3716013, adres/10.1007/978-3-540-45234-8_7, morphosys/859540, hycube/8060417}, towards larger spatial accelerators for modern analytical workloads~\cite{10.1109/ISCA52012.2021.00085, plasticine/10.1145/3140659.3080256, RIKEN/10.1145/3597031.3597055}.

Multitask execution allows kernels to co-exist on the same CGRA fabric by placing them in different partitions, thereby improving overall fabric utilization~\cite{PPA/10.1145/1669112.1669160, kong2023hardware, drips/9773269, multisky/11121915, cadas/10.1145/3793672}. However, while dynamic sharing of resources improves utilization, it also introduces fragmentation. As kernels are dynamically placed in different partitions of the array, the reconfigurable fabric becomes increasingly fragmented. Over time, the CGRA fabric may have sufficient resources in aggregate, yet still be unable to accept a new kernel because no contiguous region is large enough to satisfy its placement requirements.

\begin{table*}[t]
  \centering
  \caption{Feature comparison to related work for Virtualized Multitask CGRAs. DRIPS\cite{drips/9773269} and consequently MultiSky\cite{multisky/11121915} support a pipeline transformation mechanism. While these mechanisms resemble migration to a limited extent, they serve a different purpose altogether and therefore are not considered an overlapping feature in comparison to \workname.}
  \label{tab:Contribution Table}
  {\small
  \setlength{\tabcolsep}{4pt}
  \renewcommand{\arraystretch}{1.2}
  \begin{tabular}{c || c | c | c | c | c}
    \hline
    Work  & \makecell{Kong et al.\cite{kong2023hardware}} & \makecell{CADAS\cite{cadas/10.1145/3793672}} & \makecell{PPA\cite{PPA/10.1145/1669112.1669160}} & \makecell{DRIPS\cite{drips/9773269}~/~MultiSky\cite{multisky/11121915}} & \textit{\workname} \\ 
    \hline 
    \hline
    
    \makecell{Array Partitioning (Granularity)} & \makecell{Flexible(1D)} & Not Disclosed & \makecell{Flexible(2D)} & \makecell{Processing Element} & \makecell{Flexible(2D)} \\
    \hline
    
    \makecell{De-Fragmentation Strategy} & \xmark & \xmark & \xmark & Proactive & Reactive \\
    \hline
    
    \makecell{Live Kernel Migration} & \xmark & \xmark & \xmark & \xmark* & \cmark \\
    \hline
    \hline
  \end{tabular}
  }
  \vspace{-2ex}
\end{table*}

Therefore, making dynamic multi-tenant CGRAs practical requires not only runtime allocation, but also mechanisms that can actively recover fragmented fabric. \autoref{tab:Contribution Table} lists the most relevant works and their supported features, highlighting the gap in active support for fabric de-fragmentation.

To address this challenge, we present \workname, an end-to-end system for CGRA virtualization that enables dynamic kernel co-execution on disjoint vCGRA regions and integrates a reactive fabric de-fragmentation mechanism to resolve fabric fragmentation issues that arise due to dynamic allocation and release of spatial resources. We introduce \textcircled{1} stateless and \textcircled{2} stateful preemption-based region de-fragmentation mechanisms, that migrate running kernels across regions. Stateless trades execution progress for lower migration cost, while stateful preserves execution progress but is more complex and laborious. Stateful migration requires execution progress to be traceable, allowing a kernel to resume from its most recent snapshot after migration. To the best of our knowledge, no prior CGRA virtualization work explores this mechanism. 

Our evaluation shows that, even under heavy fragmented layouts, stateful migration consistently improves user-observed performance and makes de-fragmentation a practical mechanism for shared CGRA execution.

In summary, this paper makes the following contributions:
\begin{itemize}
    \item We present \workname, a virtualized CGRA architecture with flexible regions. A vCGRA region is the resource unit virtualized and exposed to the runtime. Elasticity allows for multiple regions to be fused to form larger ones to support more demanding applications.
    \item A solution to the fragmentation challenge identified on similar tiled architectures through different approaches of at runtime live kernel migration; stateless and stateful preemption-based migration. Stateful migration requires execution progress to be traceable, allowing a kernel to resume from its most recent snapshot.
    \item We show that live migration of kernels is an effective mechanism for recovering fragmented fabric resources and improving scheduling efficiency.
\end{itemize}

\section{Architecture Overview}

\subsection{Tiled Architecture \& Multitask Support}
\autoref{fig:Array Architecture} shows the CGRA fabric that is composed of Load/Store (LS) and Function-Compute (FC) PEs arranged on a grid and connected using a 2 mesh elasticized network. Adjacent PEs are grouped into regions that form the granularity that is virtualized and exposed to our runtime. Each region integrates a Fixed-Function-Accelerator Register File (FFA-RF) interface~\cite{leppanen2023} and a tightly coupled controller. FC PEs handle computation, while dedicated LS PEs move data between the pipeline and memory. The FFA-RF standardizes host to region interaction through a command passing interface for configuring (\texttt{CONFIGURE}), initiating (\texttt{EXECUTE}) and preempting (\texttt{HALT}) execution and snapshot control (\texttt{SNAPSHOT}). The controller is responsible for performing command translation, fine-grained control of the region's resources, and maintaining region metadata, including per-region availability, status, and identifier (\autoref{fig:Controller}).

To support multitask execution we statically partition our CGRA into $k$ homogeneous regions that form the virtualization granularity exposed to the runtime as vCGRA regions. We introduce dynamicity through virtualization, as these vCGRA regions are flexible and can be merged to form larger ones by our hypervisor when required (\autoref{fig:Architecture}). Our solution distributes configuration and control among vCGRA regions to enable per-region Dynamic Partial Reconfiguration (DPR) as some vCGRA regions may continue execution while other regions are being configured for a new application. For two or more regions to be joined they must be adjacent, constraining the resulting allocation to a rectangular shape. Currently, regions cannot be shared among multiple kernels.

\subsubsection{Load Store PEs}
These specialized LS PEs are responsible for exposing the memory hierarchy to a region and decoupling address generation from the compute pipeline. Each LS unit implements an affine address generator that streams data between the region and global memory, driven by a compact loop descriptor. In our current design, the loop descriptor supports up to three levels of nested loops, which is sufficient to cover loop nests of the kernels used in the evaluation stage. Each loop descriptor is implemented as an Address Generation Unit (AGU) which encodes base address, per-dimension stride, and iteration bounds, allowing the LS to produce regular access patterns without utilizing compute PEs for address calculation. The LS units can stream both from and to memory simultaneously by operating in full duplex mode when the available bandwidth allows it, which enables concurrent load and store traffic. 

\subsubsection{Function-Compute (FC) PEs}
Function-Compute (FC) PEs implement the compute datapath. Each FC PE consists of a local configuration memory that defines the operation to execute along with the sources of operands, and a local register file to store immediate constants and previous results. During execution, FC PEs operate as a streaming pipeline. Using a 2D mesh on chip network, they consume input tokens from adjacent PEs and produce tokens into the network for their neighbors to consume. Our FC PEs currently support a set of common 32-bit integer operations, such as addition and multiplication. Accumulation is implemented using the local register file feedback-loop. Basic predication is supported using a shadow predicate network parallel to the data network.

\begin{figure}[tb]
    \centerline{\includegraphics[width=.60\linewidth]{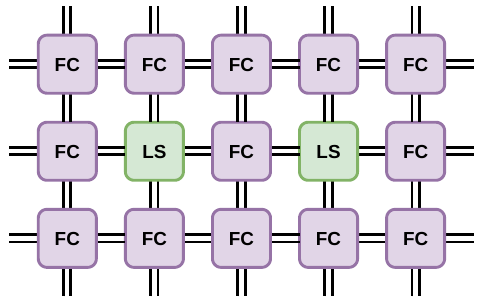}}
    \caption{Heterogeneous grid of PEs arranged on a mesh point to point network shown here for region dimensions of ($3\times5$). Regions are modular and provide full 2D flexibility.}
    \label{fig:Array Architecture}
\end{figure}

\subsubsection{Tightly Coupled Controller}
\begin{figure}[tb]
    \centerline{\includegraphics[width=.60\linewidth]{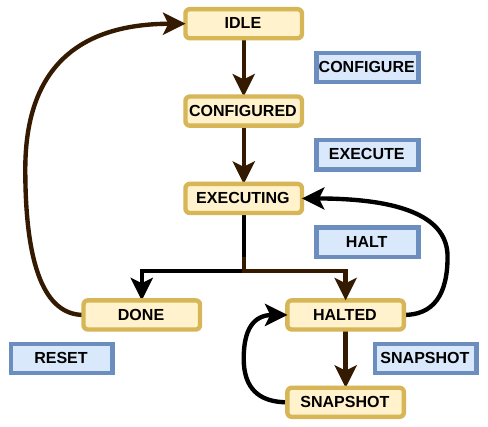}}
    \caption{FSM of our tightly coupled controller. States are shown in yellow and commands in blue. We define a minimal set of both states and commands, prioritizing utility and simplicity. A command is accepted only in its valid state, raising an \texttt{Illegal-Command} flag otherwise.}
    \label{fig:Controller}
\end{figure}

Since we aim for scalable and efficient system-level integration, we loosely couple our CGRA to the host (over PCIe) and introduce a per-region tightly-coupled controller to offload laborious fine-grained control. The controller abstracts vCGRA region management through the aforementioned command interface, reducing host-side communication latency and reporting region status back to the host. The tightly coupled controller is responsible for receiving, decoding, and executing host commands issued through the FFA-RF interface over PCIe. Specifically, we define a set of commands to support the basic operations required for fine-grained management of a vCGRA region (\autoref{fig:Controller}). To configure a vCGRA region we first issue a \texttt{CONFIGURE} command to the controller. This command takes as input the physical address of the region configuration in memory and loads the configuration to the local configuration memory of each PE. The \texttt{EXECUTE} command triggers execution of the programmed configuration. 

A \texttt{HALT} command acts as an interrupt event that freezes execution. The LS PEs stop fetching/committing any further data to memory and complete any already issued memory transactions. AGUs inside the LS PEs enter a halting state and the internal address counters stop generating new addresses. The FC PEs stop producing and consuming tokens from the network halting execution.

\begin{figure}[tb]
    \centerline{\includegraphics[width=1.0\linewidth]{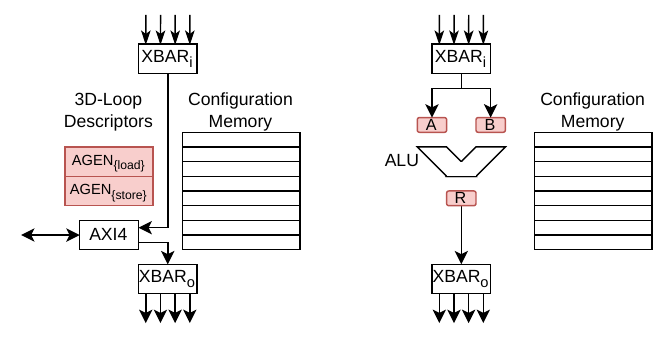}}
    \caption{Per PE type \sethlcolor{state_critical_red}\hl{state critical} elements.}
    \label{fig:Micro Architecture}
\end{figure}

To ensure correctness of execution after kernel migration it is necessary to restore the pipeline to its pre-migration state. The \texttt{SNAPSHOT} command captures a kernel's execution progress and stores it in a buffer in global memory. LS PEs expose their AGUs' progression registers to the controller, that describe the latest committed memory transaction, for load and store operations. FC PEs expose their state-critical registers that contain valid unconsumed tokens, and previous results to the controller to be included in the snapshot (\autoref{fig:Micro Architecture}).

\subsection{Shell}
\begin{figure}[tb]
    \centerline{\includegraphics[width=1.0\linewidth]{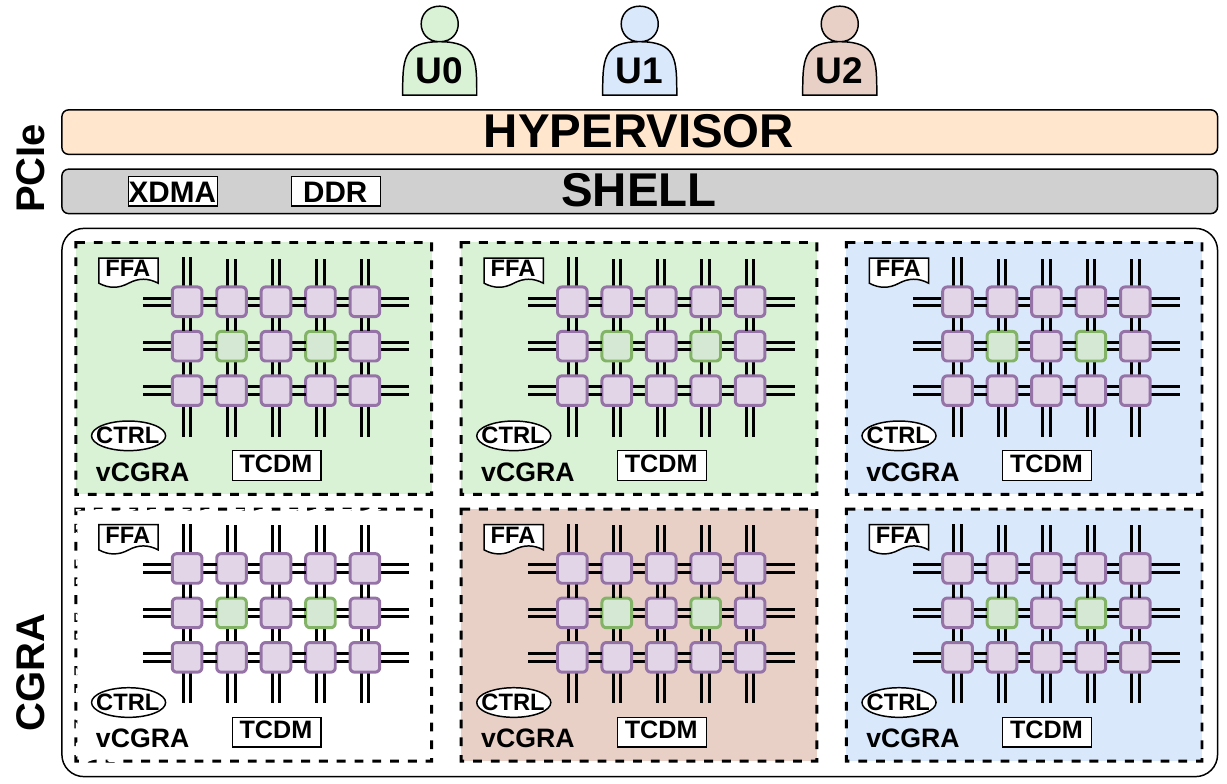}}
    \caption{System-level organization of \workname. Hypervisor receives requests from multiple users and allocates resources (color-coded). Requests are communicated over PCIe to the Shell which manages the underlying vCGRA regions. Kernels execute in parallel at different reconfigurable vCGRA regions. A kernel may occupy more than one region, in which case they are merged into one unified region.}
    \label{fig:Architecture}
\end{figure}

A shell is required to integrate our CGRA device into the system. Our shell provides a standard interface between the host and the accelerator device, and also abstracts the kernel offloading procedure for the end user. With our shell we implement a hardware interface to control and configure the accelerator's regions, realize data exchange between device and host, and most importantly, schedule kernels dynamically, and enable live kernel migration (\autoref{fig:Architecture}). Communication between host and device is realized over PCIe, by employing the XDMA core IP provided by Xilinx without modifications. Alongside the per-region tightly coupled data memories, we include a large global memory buffer within our design implemented using the on-board DDR module to store data, kernel configurations, and snapshots.  

\subsection{Hypervisor}
The hypervisor exposes the aforementioned hardware abstractions as a software interface to the end user as a runtime environment, allowing for either fine-grained control or fully opaque interaction with the device. Kernel scheduling is performed dynamically through a lookup resource map of the virtualized array. Small kernels may occupy individual regions, while larger kernels may require merging multiple adjacent regions to form wider pipelines and satisfy mapping constraints. Since all regions are uniform, placing an elaborated configuration reduces to an availability check for a sufficient set of free adjacent regions that meet the kernel compute and bandwidth requirements. This is implemented as a windowed scan of the array in an attempt to find enough contiguous regions, based on the kernel shape, to accommodate the candidate kernel. Each time the hypervisor attempts to schedule a kernel, placement may fail either due to currently insufficient compute or memory resources. On a placement failure, the hypervisor greedily checks whether fragmentation is the cause. If fragmentation is identified as the blocking factor (\autoref{sec:migration}), the array de-fragmentation procedure is initiated on a virtual image of the fabric, with the goal of reducing fragmentation while preserving the set of currently running kernels. If the resulting layout enables placement of the target kernel, the new layout is applied to the physical array by migrating the affected kernels and then placing the new kernel. Preempting a kernel flushes the configured pipeline. No further results are produced, and in-flight data are committed to state. At this stage, the hypervisor may discard the kernel state or capture it in order to reinitiate execution later. Hypervisor actions induce their own delay to the system, as pushing a kernel to the scheduler queue introduces system-side latency from serialized code and I/O.

\section{Fragmentation \& Migration Support}

\subsection{Background}\label{sec:migration}
We model a kernel as a tuple of parameters \eqref{eq:kernel tuple}, where $h_i$ and $w_i$ denote kernel height and width, the occupied area being $h_i \cdot w_i$, and $k_{id}$ being kernel id. Additional parameters may be included to carry user defined kernel metadata.

\begin{equation}
     K_i = (h_i, w_i, k_{id}, \ldots)
    \label{eq:kernel tuple}
\end{equation}

\begin{figure}[htbp]
    \centerline{\includegraphics[width=.95\linewidth]{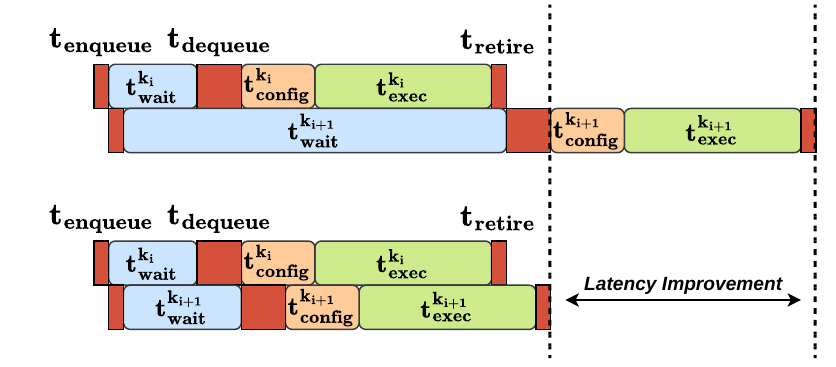}}
    \caption{A tiled multi-tenant architecture enables concurrent execution on disjoint regions and can reduce $t_{wait}$ by overlapping $t_{exec}$ of independent kernels, as illustrated. Filled red boxes denote hypervisor-induced delay. These intervals are mutually exclusive and cannot be overlapped with each other. Examples of such induced delays are, PCIe host to device communication, copy of kernel data buffer from host memory to device and reverse, dynamic resource allocation, etc.}
    \label{fig:Time Overhead Reduction}
\end{figure}

Kernel migration at runtime for CGRAs is an unexplored field with architecture specific challenges. Design space exploration of migration policies, victim selection, and compaction algorithms is out of scope for this work. Instead, our focus is on feasibility and on providing the architectural and runtime mechanisms required to enable kernel migration, including state capture and restore for stateful migration, region reconfiguration, and the control protocol needed to relocate running kernels safely. When a de-fragmentation event is triggered, we currently use a greedy compaction heuristic that defines a gravity point at the south-west of the array. We then halt all running kernels and migrate their allocated regions towards, and around, that point in order to combine free space and reduce fragmentation. This adds an overhead, since we pay the cost of migration (stateless or stateful) for all running kernels on the array. We treat fabric de-fragmentation as a corrective mechanism, that is triggered only when a kernel cannot be scheduled even though sufficient resources exist in aggregate. We estimate whether fragmentation is the blocking factor by evaluating \eqref{eq:fragmentation check}, following Septien et al.~\cite{fragmentation/fpga/1639432}, with \(A_{free}\) being the total free space available, $\alpha$ a heuristic argument and $h_i \cdot w_i$ the dimension requirements of a kernel, all counted in regions.

\begin{equation}
    A_{free} \geq \alpha \cdot h_i \cdot w_i, \qquad \alpha = 2
    \label{eq:fragmentation check}
\end{equation}

\begin{figure}[tb]
    \vspace{-5.5ex}
    \centerline{\includegraphics[width=0.80\linewidth]{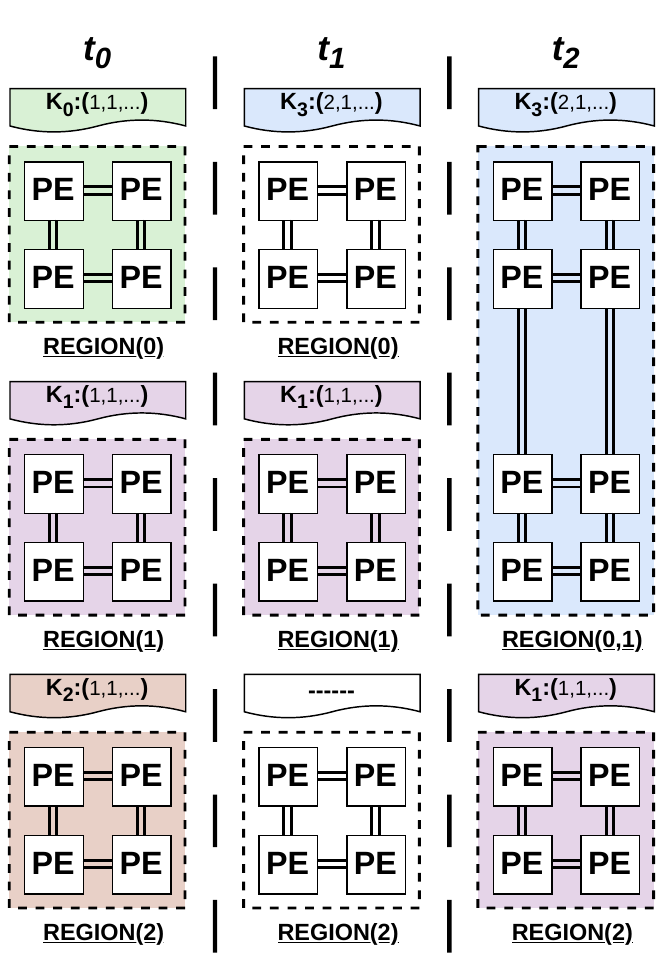}}
    \caption{Kernels may not finish execution in the same order in which they were issued the CGRA fabric may become fragmented. Initially, regions $R_0, R_1\ \text{and}\ R_2$ are occupied by kernels $K_0, K_1\ \text{and}\ K_2$. After some time $K_0\ \text{and}\ K_2$ finish execution leaving the fabric in a fragmented state prohibiting the placement of $K_3$ which requires two contiguous regions in the same column. The hypervisor decides to migrate $K_1$, de-fragmenting the fabric enabling placement of $K_3$.}
    \label{fig:Fragmentation and Migration}
\end{figure}

While virtualizing CGRAs offers the opportunity to support multi-tenancy and increase utilization of the available fabric, tiled arrays suffer from the following known challenge. As we show in \autoref{fig:Fragmentation and Migration}, kernels do not necessarily complete in the order they were issued, and since CGRAs are spatial architectures, dynamically allocating and releasing regions per issued kernel creates holes in the fabric, thus resulting in fragmentation~\cite{cadas/10.1145/3793672}. We define a hole as an idle region, or a group of idle regions, that forms a contiguous rectangle, surrounded by active regions on the sides. 

To frame our objective, \(t_{wait}^{k_i}\) denotes the time a kernel $k_i$ spends waiting in the scheduler queue, $t_{config}^{k_i}$ the time overhead of programming its configuration onto the fabric, including the host to device communication overhead (in our case PCIe), and $t_{exec}^{k_i}$ its raw execution time on the array. The end-to-end time observed by the user without accounting for system-side induced delays is given by \ref{eq:ttotal}.

\begin{equation}
    t_{total}^{k_i} = t_{wait}^{k_i} + t_{config}^{k_i} + t_{exec}^{k_i}.
    \label{eq:ttotal}
\end{equation}

Our goal is to reduce \(t_{wait}\) through spatial sharing of the fabric (\autoref{fig:Time Overhead Reduction}), rather than to improve \(t_{exec}\) of an individual kernel. In fact, \(t_{exec}\) may increase under multi-tenancy due to memory bandwidth contention and interference. In a monolithic execution model, \(t_{wait}^{k_i}\) is dominated by the execution time of previously issued kernels \eqref{eq:wait_mono}.

\begin{equation}
    t_{wait}^{k_i} = \sum^{i-1}_{j=0}(t_{config}^{k_j} + t_{exec}^{k_j}).
    \label{eq:wait_mono}
\end{equation}

\subsubsection{Stateless Migration}
Stateless Migration halts an ongoing execution, transfers the kernel configuration to a different region, and restarts the kernel. As a result, all prior execution progress is discarded, incurring lost work $t_{lost}$, and any in-flight results are not committed to memory. The corresponding migration overhead is given in \eqref{eq:statelss migration cost}. For kernels that use the tightly coupled data memory (TCDM) of a vCGRA region, the initial TCDM contents must be transferred alongside the configuration to the target vCGRA region. This operation adds a further cost of $t_{tcdm_i}$.

\begin{equation}
    t_{mig\_stateless} = t_{config} + t_{lost} + t_{tcdm_i}
    \label{eq:statelss migration cost}
\end{equation}

To improve the efficiency of Stateless Migration, we introduce a tolerance threshold that avoids migrating kernels that are near completion. The threshold provides additional flexibility by filtering out migrations whose expected benefit is limited by the small amount of remaining work. We estimate the execution progress coarsely using the ratio between the current iteration count and the total iteration count \eqref{eq:statelss migration cost threshold}, and migrate a kernel only when $c_{th} \leq f$. Setting $f = 1.0$ enforces migration for all kernels regardless of their progress.

\begin{equation}
    c_{th} = \frac{it_{now}}{it_{total}}, \qquad f \in (0,1]
    \label{eq:statelss migration cost threshold}    
\end{equation}

\subsubsection{Stateful Migration}
Low-level execution snapshots are inherently design-dependent, since each architecture defines its own state-critical elements (\autoref{fig:Micro Architecture}). To continue execution of a preempted kernel on another region, after migration, execution progress must be traceable, so, snapshot capture reads all state-critical registers within a vCGRA region. Migration therefore, induces an additional overhead $t_{state\_regs}$ of 30\%, as compared to region configuration cost in cycles. In Stateful Migration, the TCDM contents must also be restored. Rather than retrieving them from the source configuration at migration time, as in the stateless approach, stateful migration reloads the TCDM contents from the kernel's most recent snapshot \eqref{eq:stateful migration cost}, increasing the migration cost by $t_{tcdm_c}$ which may vary. 

In summary, relative to stateless migration, stateful is more expensive in the short term, but it is also the only mechanism that preserves intermediate progress. Maintaining execution progress is of major importance, not only for efficiency, but also, in some cases, for correctness. This is especially true for non-restartable tasks whose inputs are overwritten during execution; for example $Y=X+Y$.

\begin{equation}
    t_{mig\_stateful} = t_{config} + t_{state\_regs} + t_{tcdm_c}
    \label{eq:stateful migration cost}
\end{equation}

\section{Evaluation}\label{sec:evaluation}

\subsection{Methodology}
Evaluation of \workname\ is conducted using two methodologies -- each serving a different purpose. \textcircled{1} Hardware emulation evaluates the performance gain of the tiled architecture using an Alveo-U280 by integrating it in a real system. \textcircled{2} The Python model-level simulator serves as a controlled experimentation environment for high-level fragmentation benchmarking of performance gain.

\begin{enumerate}
    \item The hardware emulation method constructs an architecture with 16 vCGRAs arranged in a \(4\times4\) grid. Each vCGRA consists of a \(3\times5\) PE grid (as shown in \autoref{fig:Array Architecture}) for a total of 240 PEs. We generate workloads of 64 jobs by drawing a random mix from the selected PolyBench routines, BLAS kernels, and machine learning operators (\autoref{tab:workloads}). These benchmarks exhibit deep loop nests, high data reuse, and regular memory access patterns that match the CGRA execution model.
    
    \item We simulate a virtual image of our (\(4\times4\))-architecture with Python and generate fragmentation intensive workloads of 64 jobs using a Genetic Algorithm (GA) over the same routine pool used in hardware emulation evaluation, although increasing the variety of allocated shapes and fluctuations in problem size, for the purpose of inducing more fragmentation to the fabric.
\end{enumerate}

\begin{table}[tb]
\centering
\caption{Selected PolyBench and ML workloads used for evaluation. Using this table we generated a total of 64 jobs that randomly mix the selected kernels.}
\label{tab:workloads}
\begin{tabular}{l l l l}
\hline
Category & Kernel & Pattern & Problem Size \\
\hline
BLAS             & gemm       & 3D loop nest, MAC         & N=128  \\
BLAS             & 2mm        & chained matrix            & N=128  \\
BLAS             & mvt        & matrix-vector             & N=512  \\
Data mining      & covariance & reduction                 & N=2048 \\
Neural Networks  & relu       & map                       & N=4096 \\
BLAS             & saxpy      & vector-vector             & N=4096 \\
\hline
\end{tabular}
\vspace{-2ex}
\end{table}

Our evaluation examines the initial hypothesis presented in \autoref{fig:Time Overhead Reduction}. We measure $t_{wait}$, $t{_{config}}$, and $t_{exec}$ using \eqref{eq:perfcount1}, \eqref{eq:perfcount2} and \eqref{eq:perfcount3}. We attach a set of reference timestamps to each kernel. We define $t_{arrival}$ as the time at which a kernel enters the hypervisor's queue and $t_{scheduled}$ when the kernel is scheduled/placed on a vCGRA. Execution begins at $t_{launch}$ and completes at $t_{completed}$.

\begin{equation}
    t_{wait} = t_{scheduled} - t_{arrival}
    \label{eq:perfcount1}
\end{equation}
\begin{equation}
    t_{config} = t_{launch} - t_{scheduled}
    \label{eq:perfcount2}
\end{equation}
\begin{equation}
    t_{exec} = t_{completed} - t_{launch}
    \label{eq:perfcount3}
\end{equation}

Performance gain evaluation employs three key metrics: \textcircled{1} \textit{Makespan} captures the total time required to complete the workload, measured as the difference between the latest kernel completion and the earliest kernel arrival. \textcircled{2} \textit{Mean} turnaround time, denoted as $\overline{TAT}$, measures the mean time each kernel spends in the system from arrival to completion. \textcircled{3} $TailLatency_{95}$ captures the 95th percentile of kernel turnaround time, reflecting the latency experienced by the slowest $5\%$ of kernels.

\vspace{-2ex}

\begin{equation}
    Makespan = \max(t_{completed}) - \min(t_{arrival})
    \label{Makespan}
\end{equation}

\begin{equation}
    \overline{TAT} = \sqrt[N]{\prod_{i=1}^{N} (t^{i}_{completed} - t^{i}_{arrival})}
    \label{TurnaroundTime}
\end{equation}

\begin{equation}
    TailLatency_{95} = P95(TAT)
    \label{eq:TailLatency}
\end{equation}

\subsection{Benchmarking through Hardware Emulation}
\begin{figure}[tb]
    \centerline{\includegraphics[width=.70\linewidth]{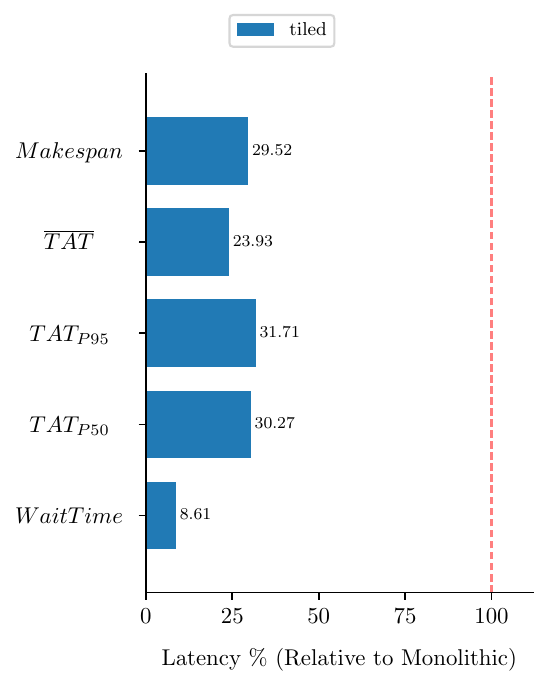}}
    \caption{($4\times4$)-architecture evaluation using hardware emulation. Mean wait time decreases by $91.39\%$ improving user experience substantially, reducing $P95$ tail latency by a factor of $68.29\%$ and specifically mean turnaround time by $76.07\%$.}
    \label{fig:Evaluation HW_0}
\end{figure}

\begin{figure}[tb]
    \centerline{\includegraphics[width=.80\linewidth]{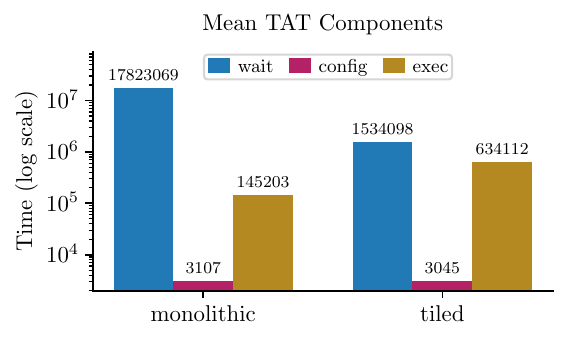}}
    \caption{Our tiled architecture decreases mean wait time by \(\times11.61\); although the mean execution time is increased by \(\times3.42\) due to memory access congestion, the total TAT is considerably improved by up to \(\times8.27\). Note that the execution time can be reduced with multi-channel memory, such as HBM. In both architectures, configuration time remains unchanged, since distributed per-region configuration maintains constant configuration overhead.}
    \label{fig:Evaluation HW_1}
\end{figure}

We evaluate latency, mean turnaround time, and makespan under tiled execution on real hardware using an Alveo-U280 FPGA board. The tiled array outperforms the monolithic baseline across all three metrics (\autoref{fig:Evaluation HW_0}) because it can schedule multiple kernels concurrently on disjoint vCGRA regions, reducing queuing delay and improving overall fabric utilization. This benefit comes at the cost of increased execution time under co-execution, as multiple kernels contend for shared memory bandwidth and introduce congestion (\autoref{fig:Evaluation HW_1}).

\subsection{Benchmarking through Simulation}
Our optimization approach explores changes in workload composition, through kernel shapes, and fluctuation in execution time that maximize fragmentation intensity. It increases the frequency of kernel shapes that are more likely to fragment the array, and stresses our dynamic architecture with out-of-order kernel completions.

\subsubsection{Dynamic Scheduling} Even for highly fragmented layouts the tiled array outperforms the monolithic baseline, which is limited to single-kernel execution across all metrics. Dynamic scheduling of multiple tasks reduces total workload $Makespan$ by $21.08\%$, $TailLatency_{P95}$ by $22.37\%$ and $\overline{TAT}$ by $17.79\%$ increasing utilization of the available fabric by placing kernels as soon as there is available space. This improves fabric utilization and reduces waiting time, resulting in better user-observed performance, which is crucial in multitenant platforms.
\subsubsection{Stateless Migration} We evaluate stateless migration using two progress thresholds, $f=1.0$, which permits migration regardless of kernel progress, and $f=0.8$. Stateless migration for $f=1.0$ forces migration for all kernels ignoring current execution progress, whereas $f=0.8$ checks if a kernel has completed more than $80\%$ of its iterations and, if so, does not relocate it.
Stateless migration recovers fragmented space and unlocks blocked placement opportunities, but its gain depends on whether the recovered space balances out the lost work of restarting the migrated kernel from the beginning. \autoref{fig:Evaluation SM} shows that stateless migration performs poorly on average because kernels must re-execute from the beginning after migration due to lack of state preservation. With $f=1.0$ stateless migration provides no benefit relative to tiled execution and in fact worsens all metrics. Stateless with $f=0.8$ provides a modest improvement, but the gain does not exceed $3\%$ across metrics.

\subsubsection{Stateful Migration} 
\begin{figure}[tb]
    \centerline{\includegraphics[width=.90\linewidth]{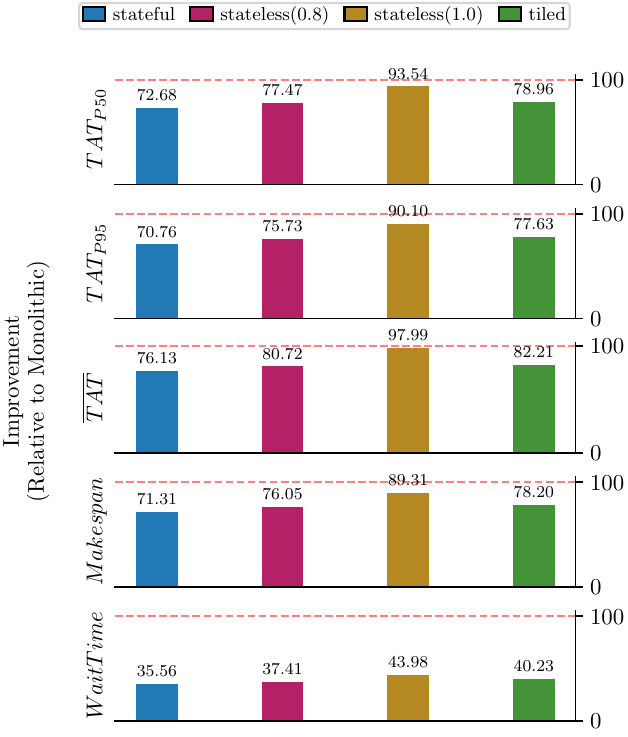}}
    \caption{($4\times4$)-architecture evaluation using selected workloads that induce fragmentation. For fairness, stateless and stateful solutions are compared under comparable fragmentation conditions, while ensuring stateful consistently performs equal to or more migrations than stateless.}
    \label{fig:Evaluation SM}
    \vspace{-1.5ex}
\end{figure}

In contrast, stateful migration preserves execution progress after kernel migration and therefore avoids the restart cost, which is the main limiting factor of stateless migration. In \autoref{fig:Evaluation SM}, stateful migration improves user-observed metrics, reducing $TailLatency_{P95}$ by $6.27\%$ and $\overline{TAT}$ by $6.08\%$ relative to tile execution as the baseline. $Makespan$ reduction shows that de-fragmentation helps recover blocked fabric and keeps the array productive instead of waiting for resources to be released naturally.

\begin{figure}[tb]
    \centerline{\includegraphics[width=0.90\linewidth]{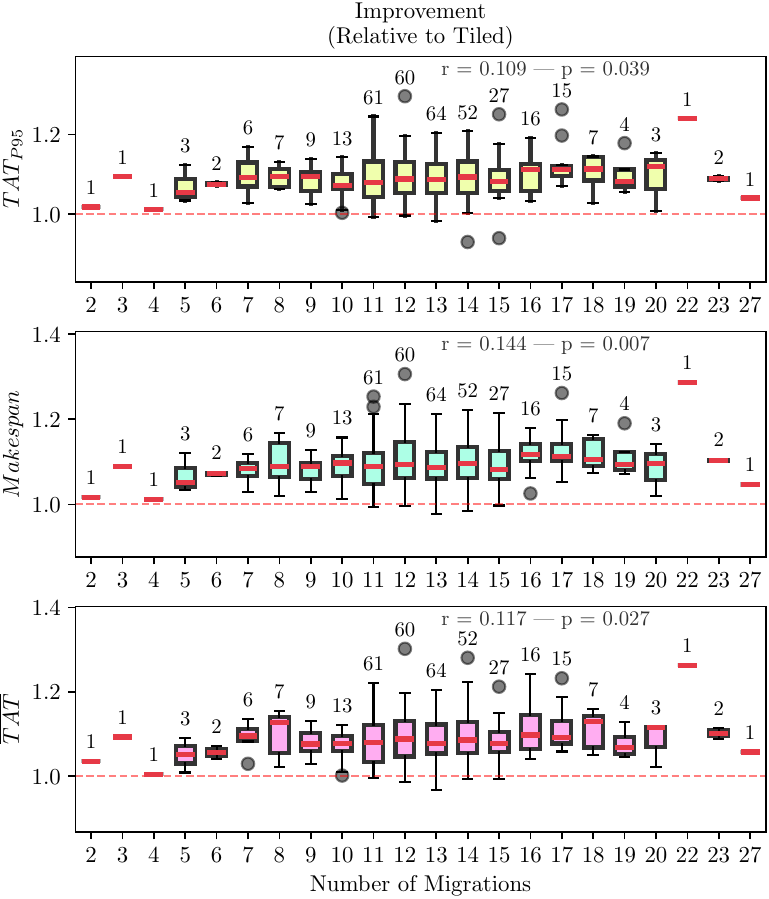}}
    \caption{($4\times4$)-architecture evaluation on the effect of the number of migrations to total performance gain with tiled as baseline. Numbers above each box denote the number of samples in each migration group; r and p indicate the strength and significance of the correlation. Overall, results suggest significant, but very weak correlation between number of migrations and performance gain.}
    \label{fig:Evaluation Stats SM}
\end{figure}

\autoref{fig:Evaluation Stats SM}, discusses the expected performance gain, as a function of the number of migrations. After thorough examination, results show a statistically significant, but very weak correlation between the number of migrations and performance gain, suggesting that, potentially, migration quality matters more than migration quantity. Stateful migration delivers better $TailLatency_{P95}$ by up to $29.60\%$ and $\overline{TAT}$ by $30.60\%$ relative to tiled execution as the baseline. High migration counts are rare, but when migration is triggered under any fragmentation conditions, stateful migration consistently improves all reported metrics.

\subsection{Resource Consumption}
Full shell synthesis for a ($4\times4$)-architecture costs 17.28\% of the total FPGA LUT resources, with total chip power measured at 3.3\,W at 150\,MHz. The architecture achieves scalability with each region consuming 1\% of total LUT resources. For 16 regions with 15 PEs per region, our platform totals to 240 PEs, which is close to the 256-PE RIKEN CGRA~\cite{RIKEN/10.1145/3597031.3597055}.

The hardware cost of virtualization and migration components (tightly coupled controller and read-back paths) introduce a LUT cost of 0.13\% per region, indicating that the cost of enabling these mechanisms is negligible relative to the overall resource footprint.

\section{Related Work}

\subsection{Multitask CGRAs}
Traditionally, scheduling multiple kernels on the same CGRA fabric is tackled at the mapping stage. Although this method is highly efficient it is inflexible and labor-intensive as it requires apriori knowledge of the tasks to be scheduled. Works such as~\cite{SchemeforMultithreadingonCGRAs/10.1145/2638558, EnablingMultithreadingonCGRAs/6047194, dpr/multikernel/10416089, FexMo/10.1145/3725843.3756019}, fall into this category, and enable multi-threaded execution on CGRAs under the Multiple Configurations Multiple Data (MCMD) execution model, as summarized in~\cite{10.1145/3357375}. A more recent work by Yang et al. presents FexMo~\cite{FexMo/10.1145/3725843.3756019}, which targets multi-task offloading on a CGRA device through multi-task static time multiplexing of Data-Flow Graphs elaborated at mapping time. The aforementioned solutions focus on static schedule transformations at compile-time for controlled sharing of the compute resources, instead of dynamically handling them at runtime. 

\subsection{CGRA Virtualization}
Virtualizing the fabric of a CGRA is a compelling direction. Prior work has explored both temporal and spatial forms of CGRA hardware virtualization. On the temporal side, Zippy and related work~\cite{zippy/1540388, zippy2/7091166} make the fabric appear larger than it physically is by time-multiplexing execution contexts. PPA~\cite{PPA/10.1145/1669112.1669160} lies between temporal and spatial virtualization, since it supports dynamic partitioning of array regions whose effective allocation changes over time according to resource availability.

\workname\ relies on spatial virtualization of hardware resources (Table~\ref{tab:Contribution Table}). To improve utilization, Kong et al.~\cite{kong2023hardware} propose a spatial sharing solution for the Amber~\cite{10258121} CGRA. However, their virtualization granularity is limited to entire CGRA columns, whereas our work, similar to \cite{PPA/10.1145/1669112.1669160}, virtualizes the array using tile-shaped groups of PEs. They also mention a bitstream relocation feature, but their solution does not support preemptive execution and therefore cannot enable live migration of running kernels, let alone stateful migration that preserves execution progress. DRIPS~\cite{drips/9773269} focuses on pipeline transformations according to input fluctuations and virtualizes the fabric at the PE level, but exhibits limited scalability and does not address the problem of fragmentation. MultiSky~\cite{multisky/11121915} aims for efficiency in frequent task creation/destruction events and improves scalability and latency through a hardware controller relative to DRIPS. Their work also discusses a proactive fragmentation-aware allocator, based on $Q$-learning, but still their solution lacks scaling capabilities, and also does not address live migration of running kernels. Lin et al.~\cite{cadas/10.1145/3793672} also study tiled CGRA execution for efficient scheduling at scale and identify fragmentation as a side effect of tiled allocation, but proposing a mitigation mechanism falls outside the scope of their work.

\subsection{FPGA Virtualization}
FPGA virtualization has received substantial attention in prior work~\cite{osdi18-khawaja, coyote/10.5555/3488766.3488822, coyotev2/10.1145/3731569.3764845, vital/0.1145/3373376.3378491, pmiliad/10.1145/3648475, nyx/10.1145/3695053.3731094, pipearch/10.1145/3418465}. These works relate to \workname\ in that they treat a reconfigurable substrate as a shared platform and expose an allocatable resource abstraction to a scheduler. However, they target FPGAs, and focus on managing bitstreams, and Partial Reconfigurable Regions (PRRs), whereas \workname\ targets CGRAs examining an entirely different level of virtualization abstraction by virtualizing the array into flexible regions that can be merged to form larger rectangular allocations. In particular, PipeArch addresses the potential of such NoC based architectures (for example CGRAs) but does not evaluate it directly. In particular, our work explores snapshot-based stateful migration which is challenging to do on FPGAs as read-back of states is not what fine grain architectures are designed for.

\section{Conclusion \& Future Work}
\workname\ is an end-to-end solution for CGRA multi-tenancy that virtualizes the array into flexible regions, and exposes them to a runtime system for dynamic allocation and scheduling. With \workname, we also explore stateless and stateful kernel migration as mechanisms to address fragmentation, a core side effect of tiled spatial sharing. Our results show that virtualizing the fabric enables substantial performance gains through parallel execution, while migration can further boost performance gains under fragmented layouts. In summary, our results indicate that stateful migration is able to consistently yield better performance across all metrics even for a high number of migrations relative to workload size. In this work, we present our observations made throughout the development process of virtualizing a CGRA suited for modern application demands and explore the feasibility, benefits and side effects of both stateless, and stateful migration for CGRAs.

\newpage
\bibliographystyle{IEEEtran}
\bibliography{refs}

\vspace{12pt}

\end{document}